\documentclass[12pt]{JHEP3}

\usepackage{graphicx}

 \preprint{YITP-05-46\\OIQP-05-10}

 \title{Law Behind Second Law of Thermodynamics
 --Unification with Cosmology--}
 
 \author{Holger B. Nielsen\\
 Niels Bohr Institute, University of Copenhagen
 17 Blegdamsvej, Copenhagen $\phi$, Denmark}

 \author{Masao Ninomiya\footnote{
 Working also at Okayama Institute for Quantum Physics, Kyoyama-cho 1-9, Okayama City 700-0015, Japan.}\\
  Yukawa Institute for Theoretical Physics, Kyoto University, Kyoto 606-8502, Japan}
 
\abstract{
 In an abstract setting of a general classical mechanical system as a
 model for the universe we set up a general formalism for a law behind the
 second law of thermodynamics, i.e. really for ``initial conditions". We
 propose a unification with the other laws by requiring similar symmetry
 and locality properties.
 }

 \keywords{Particle Cosmology, Second Law of Thermodynamics}
  \begin{document}

 \section{Introduction\label{sec1}}

 The second law of thermodynamics \cite{1,2,3} concerns, contrary to the
 typical other laws as Newton's laws in classical mechanics or say the
 Hamilton equations, the initial state conditions - or better the actual
 solution among the various possible solutions to the equations of
 motion - rather than the time development.

 It is the purpose of the present article to put forward an attempt to
 unify the second law of thermodynamics with the other laws, the time
 development laws \cite{4,5}.
 This is done by setting up a more microscopic
 formulation of this second law of thermodynamics in terms of a postulated
 ``fundamental" probability density P(``path") defined for all possible
 solutions of the equations of motion ``path". 
 A priori we would like to formulate all the laws in the language of microscopic degrees of
 freedom - these be fields or particles - rather than in terms of the
 only macroscopically understandable concept of entropy \cite{6}.

 A good unification of such a probability density $P$(``path") with the
 time development laws could be to impose the logarithm $\log{P}$(``path")
 the analogous type of symmetry and locality properties as the action $S$. 
 Even from pure esthetical considerations one could take a completely
 analogous expression as an assumption or hypothesis for $\log{P}$(``path") 
 as one writes down for the action $S$.
 If we, for instance, think of a  classical field theory
 with general relativity,
 the analogous form for the  logarithm of the ``fundamental" probability
 density would be of the form
 \begin{eqnarray}\label{1.1}
 \log P(``path")= \int
 \hat{p}(g_{\mu\nu}(x),\varphi_1(x),...\psi(x),\nonumber\\
 \partial_\rho g_{\mu\nu}(x),
 \partial_\rho\varphi_1(x),...,\partial_\rho\psi(x),)\sqrt{g}d^4x.
 \end{eqnarray}

 By imposing the same symmetry and locality and perhaps
 even ``renormalizability" restrictions on this $\log{P}$(``path") as the one on
 the action $S$ one will make to an integral expression of the 
 same form, the various coefficients on the various terms
 be in general quite different for $\log{P}$ and for the action $S$.

 In the beginning of the present article we shall keep the discussion
 so abstract that we do not even explicitly write down that we are
 concerned with a (classical) field theory with three space dimensions
 and one time dimension, but rather just describe the various modes on
 which we may expand the fields over three space as generalized
 coordinates $q_i(t)$ and their canonical conjugate momenta
 $p_i(t)$ - we imagine that e.g. spatial Fourier expansion
 coefficients of the various fields in the field theory describing the
 universe are taken as the generalized coordinates $q_i(t)$. Thus we
 must imagine that the index $i$ is a combination and ordered
 set of symbols some of which denote the type of field and the spin
 component, while there is in addition a three component wave number
 enumerating the Fourier coefficients. The $p_i(t)$ are the canonically
 conjugate momenta
 $p_i(t)=\frac{\partial{L}(\dot{q,q})}{\partial\dot{q}_i(t)}$. 
 The space coordinates which are the $p_i's$ and the
 $q_i's$
 are called phase space and has dimension $2N$, where $N$ is
 the number of $i$ values. 
 (Strictly speaking $N=\infty$ but we will often think as if it were just
 finite very large.) 
 Thus really the index $i$ here both describes which field type, which
 Lorentz or other indices and for which mode the $q_i(t)$ or $p_i(t)$
 is the expansion coefficient. For simplicity we ignore the fermion
 fields. In this way we consider the universe in a very
 abstract manner a general mechanical system, the state of which is
 described by a point in phase space and the time development. 
 It is given by the Hamilton equations
 \begin{eqnarray}
 \dot{q}_i(t)=\frac{dq_i(t)}{dt}=\frac{\partial{H}(q,p)}{\partial{p_i}},\\
 \dot{p}_i=\frac{dp_i(t)}{dt}=-\frac{\partial{H}(q,p)}{\partial{q}_i}.
 \end{eqnarray}
 A form of $\log{P}$ as given by (\ref{1.1})
 would, by making it into abstract form, become of the form $\cal P$
 \begin{equation}
 \log{P}(``path")=\int{{\cal P}(q(t),p(t))dt}.
 \end{equation}

 This general form just could be obtained by imposing a
 locality assumption in time, which leads to this expression. Then an
 assumption of time translational invariance is added to make $\cal P$ not
 depend on time explicitly - but only via the variables $q_i(t)$ and
 $p_i(t)$ ($i=1,2,...,N$ where $N$ is the number of degrees of freedom; 
 strictly speaking $N=\infty$).

 In the spirit of having in mind essentially random Hamiltonian
 $H(q,p)$ as well as random ${\cal P}(q,p)$ we just think of the phase 
 space
 as a landscape with the two types of heights $H$ and $\cal P$ which may have a priori
 randomly peaks and dips and passes and so on. In reality they
 are to obey the symmetry principles that the true laws of nature  may
 impose on them, but we may at first look for what to expect without
 putting in too many details, thus avoiding to use these symmetry
 etc. laws too much.
 
\vspace{5mm}  
 In the following section 2 
 we shall put forward how to get an idea for searching for the most
 likely the highest $\log(\left<p\right>e^s)$ - classes of path and see how such
 considerations lead quite naturally to an effective second law of
 thermodynamics. At first we only make crude suggestions. But then in
 section 3 we deliver a very general limitation on how much entropy can
 go first up and then down and a formal derivation of the second
 law of thermodynamics follows under the same reasonable approximation. But
 a further clean up is still needed to apply the second law of
 thermodynamics even for subsystems.
 In section 4 we rather shortly put a more concrete cosmological
 picture on our so far very abstract model. In section 5 we review
 shortly the various outcomes of the model behind the second law of
 thermodynamics in addition to the practical outcome of this law
 itself.
 In section 7 we conclude and present outlook especially by taking the
 lack of perfect derivation of the second law of thermodynamics as an
 extremely interesting suggestion for seeming effects of a
 foresight.

 Such effects could be strange miracle - like events seeking to prevent
 Higgs particle production, e.g. at LHC.

 \section{Seeking likely class of solutions for realization}
 \subsection{Chain of metastable high $P$ regions}

 We shall seek to get an impression of how to look for the most
 likely class of solutions if one had at one's disposal the landscape of
 $\cal P$ and $H$ over phase space.

 The most important is in fact to look for regions in phase space in which
 the system (i.e. the point describing the state of the system
 $(q(t),p(t))$) can stay around for some time, because one does not get so
 much contribution from a high $\log{P}$ contribution peak in $\cal P$, if
 the system does not remain in the region so as to get a proper
 contribution to the integral (\ref{1.1}).
 Let us in fact imagine that we investigate the landscape in the enormously
 high $(2N)$ dimensional phase space by searching for what we could call
 metastable regions in which the equation of motion solutions will stay
 an appreciable time. 
 That would be the regions where the Hamiltonian is
 near to either a local minimum or a local maximum. 
 Thus A) the partial
 derivatives of $H$ are near zero and thus small so that from Hamilton's
 equation of motion becomes slow and B) there are few of no ways
 away from it because at a minimum there will not be energy enough to
 get far away from it and at a maximum there will be too much energy to get
 away.

 Now to get a high likelihood it is crucial that $\cal P$
 contributes strongly and positively to $\log{P}$(``path")
 in the neighborhood of such an approximate maximum or minimum region
 for $H$. So we shall really only look for those maxima or minima region
 with relatively high $\cal P$-contributions to $\log{P}$. Now taking into
 consideration the huge - really infinite - number of degrees of
 freedom $N$ it seems too much to ask for free maxima or minima to
 occur so copiously that we should expect them with sufficiently high
 $\cal P$ contributions to be relevant. So we will rather think of the
 metastable regions to be like a minimum say, in $H$ in by far the
 majority of the directions in phase space, but not in truly all
 direction. So strictly speaking rather a pathlike region with only
 extremely few coordinates in which $H$ has maximum, while it is minimal in
 extremely many coordinates. In whatever way such a metastable
 regions may come about we may think of crudely ascribing them a
 ``staying time" $t$ giving the typical time at which the system will spend in
 that region following the equations of motion.
 The contribution to $\log{P}$ obtainable by having the system ``path"
 such a metastable region called $A$ is
 \begin{equation}
 \log{P}_{\mbox{\scriptsize contribution from A}}
 =
 t_{\mbox{\scriptsize stay A}}\cdot{\cal P}_A
 \end{equation}
 where ${\cal P}_A$ is the average or typical value of $\cal P$
 in this region.

 Now we are to obtain what are the most likely features of the solution
 to the equations of motion not simply to ask for that special solution
 may happen to have the very highest $\log{P}$(``path").
 The reason is that the phase space has so enormously many solutions that
 it could easily happen that there exists some class of solutions with so
 many of them that even though each of them have significantly lower
 $\log{P}$ than the very highest $\log{P}$ solutions the whole class has
 nevertheless a much higher probability to be realized than this single
 highest $\log{P}$ solution.

 When we seek the most likely \underline{class} of solutions we should
 thus also estimate the number of solution in the class proposed. 
 We would like to think of this number as an exponential of an
 entropy $e^S$. Here then the entropy for the class of solutions must be
 related to the thermodynamics entropy concept \cite{7} which we identify as the
 number of microstates - after discretization with a scale
 $U_s=\sqrt{h}$ - in a macro state. 
 The number of solutions in the class cannot be bigger than the entropy 
 at the era, the time, when the entropy is the smallest. 
 There cannot possibly be more solutions than there is
 phase space place with lowest entropy, in which time they
 are most compressed in phase space.

 Generally we may now look for a class of solutions for which it turns out
 possible to get the solutions pass through a series of metastable
 regions $A$, i.e., approximate minima or maxima for $H$ with especially
 high $p_A$ and long stay times $t_{\rm stays\ A}$. For such a class we can
 achieve a contribution to the logarithm of $P$
 \begin{equation}
 \log{P}\mid_{\rm contribution} = \mathop{\sum}_At_{\rm stay\,A} \; p_A.
 \end{equation}
 We should further take into account how big a phase space region, or
 roughly in the discretized way of thinking how many solutions, there of
 this type, this class. This number is $e^S$ where $S$ is the entropy for
 the lowest entropy one of the regions past \underline{provided} that we
 can get organized all the solution passing that metastable state to come
 through the whole series. It is not a priori guaranteed to be favorable
 to let it be so. We should in other words look for the entropies of the
 different regions defined as the logarithms $S_A$ of the phase space
 volume in the regions around the approximate and metastable min. or max.
 $A$.

 The final formula for the probability of the class becomes
 \begin{equation}
 P_{\rm class}=\exp(\mathop{\min}_A(S_A)+\mathop{\sum}_At_{\rm stayA}p_A)
 \end{equation}
 provided we can get all the $e^{\min(S_A)}$ through.

 We have been crude and worked with the approximation that, although we
 may work cosmological time unit,
 these times are still extremely short
 in any sensible units compared to numbers of the order $e^S$ where $S$ in
 an entropy or compared to the scales for $P$ variation which are supposed
 to be similar to those for $e^S$.

 \subsection{Macro variables}

 We should imagine that some combinations of the $(q_i,p_i)$'s, i.e. some
 functions of them (only very few relative to the number of degrees of
 freedom though) are what we will call ``macro variables" such as say the
 radius $a(t)$ of the universe or the stand of a piston. Although they may
 be rather few and although they also have to fulfill the equations of
motion they may be relatively so significant for what happens that we
 should imagine that there would be an effective adjustment of them taking
 place. That is to say we can imagine that some relatively few
 macro variables become adjusted in order to make the $P_{class}$ expressed
 as an exponent above become as big as possible.

 The more contributions in the sum $\mathop{\sum}_Ap_At_{stayA}$ we can
 get the higher $P_{class}$ so there will be a certain ``pressure" on the
 tuning in of the macro variables so as to several metastable regions
 being passed if possible. Since this passage will have to happen
 successively in time such a passage of several must mean significant
 development in time even in a macroscopic sense. Also if we should have coupling constants 
 as in e.g. baby universe theory be considered dynamical
 variables they should be counted as ``macro variables".

 \subsection{The era simulating big bang with bounce}

 What can we expect about the entropy at first rather the logarithms of
 the phase space volumes around the approximate minima (or maxima) of the
 Hamiltonian?
 It may depend on many details but a priori these volumes could be quite
 different even in logarithm - which is what we call the entropies -.

 Thus we can a priori expect some entropy variation with time, at least
 there seems a place for it, a first ingredient for possibly obtaining
 the second law of thermodynamics.

 The metastable region $B$ with the smallest entropy $S_B$
 is the lowest phase space $e^{S_B}$ in a series of such metastable
 regions $A$ passed by the dominating class of solutions.
 It is the one with
 a relatively low $t_{\rm stayB}$ and a very high $p_B$ because we would
 statistically expect to get ${\cal P}_B$
 which is an average over the region $B$
 huger the smaller this region i.e. the smaller $S_B$.

 Here we should think a balance among the contributions so that under
 the variation of various macro variables, $\xi$, we should have
 \begin{equation}
 0
 =\frac{\partial\log{P_{\rm class}}}{\partial\xi}
 =\frac{\partial{S}_B}{\partial\xi}
 +\mathop{\sum}_A\frac{\partial}{\partial\xi}(t_{\rm stayA}p_A)
 \end{equation}
 This should be true for all the ``initial" macro variables. 
 This fact suggests some compensations that all the
 $t_{stayA}p_A$ have got about the some orders of magnitude.
 So if ${\cal P}_c$
 large then presumably $t_{stayC}$ small relatively.

 But really it does not matter, one of the contribution will have the
 smallest volume for its metastable phase space region and we call it
 $B$.
 The era of $B$ we then take as a middle point in time $t_o$ which 
 by additive shift of the time axis we can put to zero $t_0=0$.

 We shall now suggest the interpretation that in usual cosmology we ignore
 the times on the one side of this time $t_0=t_B$ for the staying around
 the lowest entropy region $B$ and shift the sign convention so that we
 live after this era of $B$. The era before $B$ we ignore as ``pre-big bang era" 
 and not taking it seriously. Now we are in somewhat
 better setting to obtain the second law of thermodynamics
 because we have made the lowest entropy metastable state in the series.  
 Because it becomes a first era taken seriously.
 With the lowest $S=S_B$ to begin with we have
 at least better chance for increasing away.

 \section{Derivation of second law of thermodynamics}\label{sec3}

 To really get a formal argument that the entropy has to increase all
 the time we shall make use of a little lemma restricting strongly the
 possibility for a too big maximum in entropy as a
 function of time. Such a theorem is only of interest when we
 have a discussion without the second law of thermodynamics, because
 the latter not only restricts maxima but also totally forbids them in as far
 as $S(t)$ becomes monotonically increasing and thus cannot have maxima
 at all.

 We have indeed a theorem alone from the equations of motion assumed to
 be not fine tuned saying:
 If a system - with random Hamiltonian - has in time the successive
 entropies considered phase space volumes relative to a cut off unit
 $u^{2N}_s=h^N$ say, $S_1, S_2$ and $S_3$ under a ``reasonable time
 interval", then we have 
 \begin{equation}
 S_2{\leq}S_1+S_3.
 \end{equation}
 The requirement of the ``reasonable time scale" means
 that the time interval involved is so small that the number of cells of
 size $u_s^{2N}$ past during it is exceedingly small compared to
 exponentiated entropy numbers $e^s_1,e^s_2,e^s_3$.
 
 This theorem is proven by considering that the number of cells in the
 macro state with the middle time entropy $S_2$ which will reach the
 $S_3$ entropy macro state can at most be $e^{S_3}$, apart from an unimportant reasonable
 factor which makes up a fraction $e^{S_3-S_2}$ of all the states there. 
 Statistically the fraction of the $e^{S_1}$ cells into which
 the cells in the first macro state - the one with entropy $S_1$ - can come,
 which can go on to the third macro state is thus $e^{S_3-S_2}$. 
 This means
 that $e^S_1$. $e^{S_3-S_2}$ cells come through all three macro states, 
 but that is more than one for $S_1+S_3-S_2\geq0$,which is just the
 condition we were to prove. This theorem seems essentially to disagree
 with the two entropies idea propose by one of us (H.B.N.) and L.E.Rugh \cite{8}
 following Hartle and Hawking \cite{9}.
 
 Combining this theorem with the hypothesis that the minimal entropy $S_B$ 
 metastable region passed in the development is very small compared to the
 later ones (or better the ones with higher $|t-t_0|$ i.e. further out)
 at least some of them, we get the second law of thermodynamics.
 Indeed denoting the entropies at times $t$ and $t'$ by $S(t)$ and
 $S(t')$ and assuming $t'$ further out than $t$ i.e. $|t'-t_0|>|t-t_0|$,
 and on the same side of $t_0$ we can apply our theorem on the three
 successive times $t_0<t<t'$ or if $t$ and $t'$ are on the other side
 $t'<t<t_0$. The theorem then gives the condition
 \begin{equation}
 S(t){\leq}S(t_0)+S(t')=S_B+S(t').
 \end{equation}
 With the extra hypothesis that $S_B$ is exceedingly small we get
 \begin{equation}
 S(t) \leq S(t'),
 \end{equation}
 which is the second law of thermodynamics and we are through deriving it.

 How likely is it, however, that $S_B$ is indeed so small?

 It is not so unrealistic that the metastable region $B$ with the smallest
 entropy is really not metastable at all, but rather only one cut off cell
 of volume $u_2^{2N}=h^N$ with an extremely high $\cal P$ value so that 
 even
 from the passage of about one cell there comes a significant contribution
 to $\int{{\cal P} dt}$ from it.
 If ${\cal P}$ has enough variation at small distances in phase space such 
 a happening is not unlikely.

 \section{How to see our model more concretely in cosmology}\label{sec4}

 How does this abstract picture of a universe passing through a series
 of or at least some metastable macro states with especially high
 $p$-contribution match with the phenomenological picture of the universe
 development? One thing which is usually assumed in realistic field
 theoretical model - on grounds which are mainly phenomenological though - is
 that there is a bottom in the Hamiltonian. That feature we can now
 consider among the predictions of the present model behind the second law
 of thermodynamics.

 Perhaps one does not always appreciate how much could have gone wrong
 with this bottom property because one would tend to forget that also a wrong
 sign on the coefficients of the kinetic terms in the Lagrangian density
 for a boson field $\frac{1}{2}(\partial_u\varphi)^2+m^2\varphi^2$ easily
 could have spoiled the bottom. 
 The fact that different gauge fields in nature indeed have the same sign on the kinetic 
 terms is non-trivial and a remarkable property for a non-simple Lie algebra or rather Lie group theory. 
 Concerning simple Lie groups a lot of them should be discarded on the
 grounds of such a positivity requirement for the Hamiltonian, but Nature
 has chosen just the compact Lie groups suitable for having a bottom!

 It should be remarked however that with respect to gravity we cannot say
 that we have a bottom in the energy density or Hamiltonian density.
 Strictly speaking it is a complicated story even to define what we should
 identify as energy \cite{11}, \cite{12} to get a sensible one. A slightly different
 point of view on energy in general relativity is to consider some of
 Einstein equations or, if we think quantum mechanically (what we do not take into account of 
 in the present article), the Wheeler-de Witt equation that as a
 constraint total energy density gravitational plus matter energy density
 is zero. This point of view is analogous to absorbing the $divD$ in the
 Maxwell equation $div\vec{D}=\rho$ into the charge density and declaring
 that this Maxwell equation is a constraint that tells that the
 ``\underline{total} charge" $\rho-div\vec{D}$ is zero,
 considering -$div\vec{D}$ a sort of ``electromagnetic extra charge"
 analogous to the ``gravitation energy density." With latter point of view
 the gravitational energy is just equal with opposite sign to the matter
 energy and we have \underline{no bottom} in it.

 Now, however, since gravity is under present circumstances a weak
 interaction the bottomlessness clue to gravity may not severely
 threaten the metastability. Realistically we might think of some
 global or perhaps for limited region crunch of the universe. In such a
 case so strong gravity fields may appear and the bottom is (effectively)
 lost. In the light of the facts of that gravity spoils a perfect bottom but that
 there is a good approximation a bottom in the matter part of
 Hamiltonian it seems that our present universe fits wonderfully to a
 metastable region. The stability may also correspond to an extremely
 long stay time $t_{stay}$. At least we have stayed in a similar state for
 of the order of 13 milliard years.

 The discussion in the last section suggested that the middle era around the
 region $B$ with the smallest entropy $S_B$ - taken in our interpretation
 as the earliest era - be rather unstable than truly metastable. 
 It should be a state with an especially high $p$
 selected, without too much caring for the metastability. 
 This could mean in the General Relativity analogous
 ansatz form that a combination of scalar field
 values should be used with the ``potential part" of $\hat{\cal P}$, 
 especially probability favorable. 
 That is to say that the scalar fields should be near
 a maximum in the latter potential part of $\hat{\cal P}$.
 This means that some value combination should be used all over the universe 
 if, as the ansatz suggests, we assume the $\hat{\cal P}$ translational invariant 
 in both space and time.
 
 This leads to an inflation period \cite{13} of the translational invariant type
 rather than of the chaotic type.
 In this way we can claim, within only natural assumptions,
 to predict homogeneous inflation.
 Although it will clearly be highly favorable to
 probability to keep the scalar fields extremely slowly rolling down
 the potential, it is not so obvious if it can indeed be organized by
 setting variables in the ``initial state" -
 because in fact the rolling rate is determined by the equations of
 motion and the mechanical potential $V_{eff} (\varphi_1,...)$ in the
 Lagrangian $L$. Unless the manipulation so as to optimize
 $\left<P\right>e^S$ as
 discussed above can be extended to become also a manipulation of the
 coupling constants determining the inflation effective potential the
 optimizing $\left<P\right>e^s$ will, however, not be able to adjust the
 potential form as is needed to ensure slow rolling. 
 But whatever this
 optimization has to organize will be organized to
 obtain the slow rolling. So if there is any chance our model goes in the
 direction of doing what it can solve the slow rolling problem.

 It should be stressed as an interesting and presumably unexpected
 feature that our model does not favor a true big bang singularity,
 but rather that as just discussed above an as slowly rolling as it can
 organize it is singularity free inflation even at the midpoint $t=t_0$ of
 time, in the middle of era $B$ in the above notation. 
 Actually our picture is a de Sitter model with positive spatial curvature 
 - to keep the universe finite and give a finite positive turning point radius
 - which before $t_0$ contracts, reach a finite value
 at $t_0$ and then expands  first slowly and faster and faster
 dominated by the effective cosmological constant from the effective
 potential around the scalar fields close to the very
 $\hat{\cal P}$-favored value-combination. 
 Our scenario is thus smoother
 than the usual one with true big bang singularities. The very highest
 Planck temperatures may never have been realized in our scheme!

 \section{Some results and predictions}\label{sec5}

 We have already written about several outcomes of the model in the foregoing sections 
 in addition to the derivation of the effective raising of
 the entropy on our side of the middle, our region of the $t$-axis.
 We suggested that entropy raises as one goes along the
 time axis away from ``the middle point" $t_0$ (called $0$) roughly
 characterized as the era $B$ with the smallest entropy $S_B$.
 In fact we found in addition the fact that we should have an approximate bottom
 in the Hamiltonian density so as to ensure metastability to make the most
 possible use of high $\cal P$ around the state of the universe
 as was actually found.
 Further we managed as suggestive outcome, from longer time scales,
 to see effective big bang although we do not have
 a genuine singularity but rather a smooth de Sitter scenario;
 first contracting - but with inverted second law of thermodynamics -
 and then bouncing and expanding, inflating.

 As we have already touched upon a bit it would be very tempting to talk in 
 this language about the $P$ optimization as if it were a
 person with will and the $P$-governing of the solution
 if it could also influence the coupling constants in the Lagrangian. 
 That could easily turn out to be helpful to get a
 higher likelihood chance of tracks.
 In some theories like baby universe theory, \cite{14}-\cite{17}
 such influence can come as a consequence of
 what is effectively outside in our phenomenological universe theories
 - the baby universes -
 while in others you can imagine some effective dependence
 of the effective couplings on flat directions or the like that. 
 Also in formulating the fundamental theory one could just choose to include the
 couplings, mass parameters and kinetic term coefficients $Z_i$ under
 the dynamical variables so that they are allowed to have the
 possibility of having many - all real values - as possible values.
 In this way the $P$-probability machinery could come to determine
 also the values of the coupling constants etc...

 Whatever the excuse or the mechanism behind we already alluded to
 the fact that allows the parameters in the inflation potential to be
 ``dynamical" and thus to be allowed to participate in maximizing
 $\left<P\right>e^S$, it
 would make them tune in favor of a more slow rolling during inflation.
 This might
 solve the problem of the slow rolling problem requiring such an unnatural and
 strangely tuned potential. It was indeed tuned according to our model.
 Perhaps there is even some $P$-benefit by having non-zero but some
 small density of matter in the universe at a certain time!

 May be the best is to suggest that by $50\%$ chance at least there is a
 making bigger universe radius $a(t)$ more unlikely than smaller ones.
 Then a balance may appear. 
 A most interesting result that can suggestively come out once the
 coupling constants etc. are taken to be ``dynamical" (meaning that
 take different values) is the solution of the famous cosmological
 constant problem: Why is the dressed or effective cosmological constant 
 so extremely small in Planck units? Assuming that there is some term
 in $\cal P$ making it favorable to keep $a(t)$, the universe radius not 
 too big we may argue for a slow expansion of the universe radius. 
 Indeed it is very important for preventing from crunching or collapse 
 via black holes that the matter and radiation density be
 sufficiently low so that big local crunches do not take place. 
 So to keep it proper metastable there is a need for big size of the universe
 so as to get the density down to prevent from crunches or collapses via
 black holes. However these crunch dangers fall to exceedingly low
 probabilities already with today's densities. So further reduction of
 the density is not needed for exceedingly long time already. Thus with
 the assumed term seeking to keep the universe radius from going
 to infinity is there - at least $50\%$ chance - then the radius will
 settle to grow slower and slower. Since also density must be low in
 the assumptions of radius itself already large one can simply,
 from the Friedmann's equation, see that a small cosmological constant $\Lambda$ is
 needed. If thus $\Lambda$ is adjustable it will be put
 small! This is a solution to the cosmological constant problem 
 in our theory presented in this article. 
 It should, however, be admitted that the best attempts to solve the
 cosmological constant problem in literatures \cite{18}, \cite{19} already went this way of
 making it effectively dynamical one way or the other and then the
 most important part of solving the cosmological constant problem is over. 
 
 We must, however, stress that a complete solution of this
 famous problem \cite{20} must also contain what then determines the value of
 such a ``dynamical" cosmological constant just to be small. In
 principle if it has become ``dynamical" dependent on dynamical
 variables, on initial conditions so to speak, one needs a ``theory of
 initial conditions" to settle the value to be chosen. Here it
 should be mentioned that since the cosmological constant $\Lambda$ -
 or any other coupling constant or mass parameter (=bare mass)
 is constant as function of space and time, there is no
 special meaning in assigning it them to any special moment of time,
 like there is for other dynamical variables. It is therefore a priori
 not clear how one should specify coupling constants as ``initial
 conditions" unless one has as we proposed in the present article a
 more detailed model for what ``initial conditions" shall be. 
 
 In this sense we can claim that this sort of behind the second law of
 thermodynamics law is needed whenever a solution to the cosmological
 constant problem is sought by making the cosmological constant
 $\Lambda$ a ``dynamical" variable, a part of initial condition. On the
 other hand one can hardly hope to solve the cosmological constant
 problem without making the cosmological constant in some way or the
 other depend on the happenings in universe. So it almost has to be
 ``dynamical" and our type of model is thus needed!

 Let us finally mention that we hope in a soon to come article
 \cite{21} to derive from the present model the more detailed result of
 what we have called the multiple point principle \cite{22} and one of
 us (H.B.N.) and his collaborators worked a lot on from a more phenomenological
 point of view. This multiple point principle which we seem
 to be able to derive from the model says that there are many minima in
 the scalar field effective potential that are all going to be fine tuned
 into having almost the same cosmological constant or potential
 height, in the present model essentially zero, from Planck scale point
 of view.

\section{How the dependence on the future got diminished}

The arguments given above for the second law of thermodynamics 
were not very convincing, because the solution probability density 
{\em does} depend on what the track does at all times, also on times 
later than the time at which one performs some test of the second law of 
thermodynamics. 
Therefore one might imagine that some compromise could come
up and some adjustment take place to reach from the moment view point of 
future high $log P$-contribution. That is what would look like pre-arrangements.
The question really is why such pre-arrangements are not much more common
experimentally.

There is indeed an argument that such selection of the solution influenced by the future 
will be suppressed. 
In fact we shall see by arguing that 
under the conditions for which we already argued to occur in the asymptotics
the probability density will be almost constant as a function of the 
behavior of the solution $``path"$ in this future. 

Then we argued for the asymptotic situation to have the properties 
- already fulfilled in the present time of the universe development -:

\begin{enumerate}

\item Essentially only \underline{massless} particles
\item Lorentz invariance 
\item Low density $\Rightarrow$ little interactions

\end{enumerate}

In order to estimate that $P$ is indeed practically constant as a 
function of the behavior of the solution in such a part of the time,
instead of treating the universe by field theory as a classical approximation, 
we shall rather think of particles (quanta). 
Of course at the end we have the belief that the theory
should be treated quantum mechanically, but we can approximate a quantum field 
theory by a classical theory in different ways; we can either treat 
it as a theory of classical fields or as a theory of classical particles.
These two are not the same approximation and in general only good 
approximations under different conditions.

The crux of the matter is that if we should write an analogue to the field 
theory expression
for the particle classical approximation, then the free particle 
contribution to the 
$ \log P(``path")$ would have to be a sum over all particles of 
integrals along the time track of these particles in space time. The quantity 
to be integrated along the time tracks can only depend on the internal 
quantum numbers and polarization of the particle in question, while rotational 
invariance grounds it cannot really depend on polarization. 
The expression would be of the form
\begin{equation}
\log P(``path") = \sum_{i} k_i \tau_i + ``interaction ~~contributions"
\end{equation}  
where the sum runs over the individual particles $i$, and $\tau_i$ is the 
eigentime for the particle from it were created till it gets annihilated,
and $k_i$ a number characteristic of its species and internal quantum 
setting.
  
The point is that if there remain mainly massless particles then 
their eigentimes are always {\em zero.} So the contributions from massless 
particles are indeed negligible except for the contributions when they meet 
and can interact. But if interaction is also negligible in the future times 
then even such interaction caused contributions will be small.
Under these condition, $P$ is practically constant as a function of the 
behavior of the solution in era like the  asymptotic one!

Provided the future is of this nature then we deduce the following statement:

No adjustment to future, i.e. no hand-of-God-effects.

What we see with $2^{nd}$ law of thermodynamics, i.e, no regular future, 
is thus just what comes out when the future due to dominance of
massless particles is not influential on the choice of the solution $``path"$
thus leaving it for the inflation time physics to decide. 

The formal derivation, if you have physical conditions are o.k.,
may sound like this:

Why is $S$ in the short time middle state appreciably lower than outside? 
May be we can argue in the following way: There will likely be enormous peaks in
   $p(t)$ (or $P$) so that even a short stay in time around such peak
   could be of significance to bring over all likelihood up.\\
But most likely a high peak is a narrow peak - as we would expect 
for a reasonable ansatz for a ``random" function $p$- 
the more you average, then less outrageous
   average. Thus to get use of a high peak presumably a very low
   entropy only can be allowed.
This means $S$ must at least increase away on average from this
   special time $t_0$.
A truly useful $2^{nd}$ law of thermodynamics has to work for
   sub-systems separately.
Recall our theorem against entropy in middle more than twice average of side times.

Could we extend the theorem also to work more locally or rather for subsystems? 
Of course yes.

 \section{Conclusion and Outlook}\label{sec6}

 We have, in the present article, put up a very general formalism for how
 one way formulate a law behind the second law of thermodynamics, or we
 could say more generally a law for which actual solution to the equations
 of motion are the more likely. 
 In guessing such a law it is natural to
 seek to impose on the probability $P$ (``path") the symmetries of the
 other - i.e. time development - laws, even time reversal invariance and
 locality in time (and space), but that would a priori look like causing
 problems in obtaining the second law of thermodynamics. 
 
 We therefore
 consider it a quite remarkable result that we, in spite of this
 a priori unpromising outlook, in fact argue for the second law of
 thermodynamics to come out, even without to specific assumptions about
 the probability density functional $P$ in our ``model". 
 Here ``model",
 really the formalism is so general that there is only very little content,
 until one helps by adding for a speculations about properties of
 the functional $P$! Nevertheless we got second law of thermodynamics with
 only very mild such further speculative assumptions, essentially we got
 it for  ``random" functional $P$ with some locality in time and time
 translational invariance.

 The most interesting is that we only obtain it approximately and 
 in practice, but not totally exactly. Especially since in our model the
 probability density functional also depends on the behavior of the
 solution to the equations of motion in the {future}, the model is a
 priori likely to predict effects of seemingly foresight or the
 Hand-of-God. Indeed the prediction of an effective bottom in the
 Hamiltonian in practice and even of a small cosmological constant, and
 the multiple point principle \cite{22} (see in section 5 at the end of the present article) 
 may be considered such foresight arrangements. In this sense we have already in
 this article discussed how some such foresight effects are indeed
 successful predictions for our model. Since also an approximate - and
 good enough - big bang phenomenology with homogeneous inflation
 comes out as a prediction we can consider the predictions very
 successful. There is though in our picture no big bang singularity \cite{23}.

 One could consider any thought of foresight effect meaning that something
 in the present is adjusted to some special or simple coincidence in
 future as a violation of the second law of thermodynamics. If some
 features today lead to a special state tomorrow it means that ordering
 some times in some cases increase as time goes on which is just against
 the second law which says that order gets less and less with time. Our
 model is a priori filled with such foresight effects and our formal proof
 of second law of thermodynamics only means that we crude by get rid of
 them. So we do \underline{not} predict that such miraculous effects are
 totally excluded in our model. A support our model over a more pure and
 exact second law of thermodynamics would precisely be if such prearranged
 Hand-of-God effect shows up convincingly. We claimed that large universe
 radius low density bottom in Hamiltonian are already in our model of the
 Hand-of-God type.
 
 \begin{acknowledgments}
 The work is supported by Grand-in-aids for Scientific Research on Priority Areas, 
 Number of Area 763 ``Dynamics of Strings and Fields", 
 from the Ministry of Education of Culture, Sports, Science and Technology, Japan.

 \end{acknowledgments}

 \end{document}